\documentclass[5p]{elsarticle}

\usepackage{graphicx}% Include figure files
\usepackage{dcolumn}% Align table columns on decimal point
\usepackage{amssymb}% outlined letters
\usepackage{amsmath,amssymb,amsfonts}

\begin{document}

%Macros
\newcommand{\Eq}[1]{\mbox{Eq. (\ref{eqn:#1})}}
\newcommand{\Fig}[1]{\mbox{Fig. \ref{fig:#1}}}
\newcommand{\Sec}[1]{\mbox{Sec. \ref{sec:#1}}}

\newcommand{\be}{\begin{equation}}
\newcommand{\ee}{\end{equation}}
\newcommand{\bea}{\begin{eqnarray}}
\newcommand{\eea}{\end{eqnarray}}
\newcommand{\vu}{{\mathbf u}}
\newcommand{\ve}{{\mathbf e}}

%=====================================================================
%=====================================================================
%=====================================================================

\title{Thermal dimension of quantum spacetime}

\author[roma]{Giovanni Amelino-Camelia}
%\affiliation{\addressRoma}
\author[imp,bol]{Francesco Brighenti}
%\affiliation{\addressImperial} \affiliation{\addressBologna}
\author[imp]{Giulia Gubitosi}
%\affiliation{\addressImperial}
\author[roma]{Grasiele Santos}
%\affiliation{\addressRoma}

\address[roma]{Dipartimento di Fisica, Universit\`a ``La Sapienza''€œ
and Sez. Roma1 INFN, P.le A. Moro 2, 00185 Roma, Italia}
\address[imp]{Theoretical Physics, Blackett Laboratory, Imperial College, London, SW7 2BZ, United Kingdom}
\address[bol]{Dipartimento di Fisica e Astronomia dell'Universit\`a di Bologna and
Sez. Bologna INFN, Via Irnerio 46, 40126 Bologna, Italy}

\date{\today}

\begin{abstract}
Recent results suggest that a crucial crossroad for quantum gravity is the characterization of the effective dimension of spacetime at short distances, where quantum properties of spacetime become significant. This is relevant
in particular for various scenarios of ``dynamical dimensional reduction"
which have been discussed in the literature. We are here concerned with the fact that the related research effort has been based
mostly on analyses of the ``spectral dimension",
 which involves
 an unphysical Euclideanization of spacetime and is highly sensitive to the off-shell
  properties of a theory.
 As here shown, different formulations of the same physical theory can have wildly different spectral dimension.
 We propose that dynamical dimensional reduction should be described in terms of the ``thermal dimension" which we here introduce, a notion that only depends on the physical content of the theory.
 We analyze a few models with dynamical reduction both of the spectral dimension and of our thermal dimension,
 finding in particular some cases where thermal and spectral dimension agree, but also some cases
 where the spectral dimension has puzzling properties while the thermal dimension gives a different and meaningful picture.
 \end{abstract}

\maketitle

%=====================================================================
%=====================================================================
%=====================================================================

%\section{Introduction}

\section{Introduction.}

There are several alternative approaches to the study of the quantum-gravity problem, with formalizations and
physical pictures that are significantly different, in most cases offering very few opportunities to compare
predictions between one approach and another. As a result, there is strong interest for the few features which
are found to arise in several alternative
models. A common expectation is that at short distances the classical picture of spacetime as a Riemannian geometry
should be replaced by some new ``quantum" geometry. The alternative pictures of quantum spacetime
appear to be rather different, but over the last decade it became clear that some of these have in common the mechanism of ``dynamical dimensional reduction":
the familiar four-dimensional classical picture of spacetime in the IR (``infrared", {\it i.e.} for probes of wavelength much longer than the Planck length) is replaced
by a quantum picture with an effective number of spacetime dimensions smaller than four
in the UV (``ultraviolet", {\it i.e.} for probes of wavelength comparable to the Planck length).
These exciting recent developments face the challenge that the standard concept of dimension of a spacetime, the ``Hausdorff dimension",
is inapplicable to a quantum spacetime \cite{Wheater:1998jb, Carlip:2009kf}, and therefore one must rely on some suitable new concept. This challenge has been handled so far mostly\footnote{{Other candidates for the characterization of the dimension of a quantum spacetime have been proposed in Refs. \cite{Carlip:2009kf, Calcagni:2012rm, Reuter:2011ah, Niedermaier:2006ns, Padmanabhan:2015vma}.}} by resorting to  the notion of ``spectral dimension",
whose key ingredient is the (modified) d'Alembertian  of the theory\footnote{There are cases, such as in Causal Dynamical Triangulations, where the d'Alembertian of the theory is not known, but it is possible to calculate the spectral dimension with other techniques. It has been established \cite{Sotiriou:2011aa} that in these cases it is then possible to reconstruct the d'Alembertian.}
 and for classical flat spacetimes
 reproduces the Hausdorff dimension. It was in terms of the spectral dimension that dynamical dimensional reduction was described for several approaches
 to the quantum-gravity problem, including the approach based
 on Causal Dynamical Triangulations \cite{Ambjorn:2005db}, the Asymptotic-Safety approach \cite{Litim:2003vp},
  Horava-Lifshitz gravity \cite{Horava:2009if}, the Causal-Sets approach \cite {Belenchia:2015aia},
  Loop Quantum Gravity \cite{modesto, oriti},
  Spacetime Noncommutativity \cite{Benedetti:2008gu} and theories with Planck-scale curvature of momentum space \cite{DSRRSD, Arzano:2014jua}.

It is for us cause of concern that so much of our intuition about the quantum-gravity realm is being
attached to analyses based on the spectral dimension, which is not a physical characterization of a theory.
For such precious cases where a feature is found in many approaches to the quantum-gravity problem,
and therefore might be a ``true feature" of the quantum-gravity realm, we should ask for no less than
a fully physical characterization.
It is well known that the spectral dimension provides a valuable characterization
 of properties of classical Riemannian geometries \cite{Benedetti:2008gu, Vassilevich:2003xt}, but its proposed applicability to the description
 of the dimension of a quantum spacetime involves some adaptations, and, as we shall here see, these adaptations are responsible
 for some of its inadequacies. In the study of quantum spacetimes
the spectral dimension is the effective dimension 'seen' by a fictitious diffusion process governed
 by the {\underline{Euclidean version}} of the d'Alembertian. The UV value of the spectral dimension, $d_{S}(0)$, is then formally computed via:
\be
d_{S}(0)\equiv -2 \lim_{s\rightarrow 0}\frac{d\ln P(s)}{d\ln(s)}\, ,
\label{dsof0}
\ee
where $P(s)$  is the average return probability of the diffusion process and $s$ is the fictitious diffusion time.
When the IR Hausdorff dimension of spacetime is $D+1$, and the Euclidean d'Alembertian of the theory is represented on momentum space as $\Omega^{E}(E,p)$,
the return probability is given by\footnote{ Our thesis here is that even if (\ref{eq:returnP}) did describe the return probability
(as usually assumed)
still the spectral dimension would be unsatisfactory. It is interesting however that, as stressed in Ref. \cite{ASTRID},
the interpretation
of (\ref{eq:returnP}) as return probability is not always applicable.}
\be
P(s) \propto \int dE \,dp \, p^{D-1} \,
e^{-s\,\Omega^{E}(E,p)}\, .\label{eq:returnP}
\ee
The fact that the Euclidean version of the d'Alembertian intervenes is of course cause of concern for us \footnote{{Concerns for the
 Euclideanization involved were also raised in Ref.\cite{Eichhorn:2013ova}, within a study concerning the causal-set approach.
  Ref.\cite{Eichhorn:2013ova} proposed a possible redefinition of the spectral dimension suitable for including Lorentzian signature  and
   found that it gave different results with respect to the standard (Euclideanized) spectral dimension. }}. It is in fact well known
that the Euclidean version of a quantum-gravity model can be profoundly different from the original model in Lorentzian spacetime
(see, {\it e.g.}, Ref.\cite{Carlip:2015mra}).
Moreover, evidently
 in (\ref{eq:returnP}) an important role is played by off-shell modes, a role so important that, as we shall here show,
one can obtain wildly different values for the spectral dimension for different formulations of the same physical theory
(cases where the formulations coincide on-shell but are different off-shell). We are also concerned by the fact that evidently
the $P(s)$ of (\ref{eq:returnP}) is invariant under active diffeomorphisms on momentum space
(an active diffeomorphism on momentum space amounts to an irrelevant change
of integration variable for $P(s)$). Since an active diffeomorphism can map a given physical theory into a very different one
(also see here below), we believe that this degeneracy of the spectral dimension is worrisome.

While these concerns are, in our appreciation, very serious, we do acknowledge that several analyses centered on the spectral dimension
give rather meaningful results. Therefore we are here guided by the idea that it is necessary to replace the spectral dimension
with some other fully physical notion of dimensionality of a quantum spacetime, with the requirement  that in most cases
the new notion should agree with the spectral dimension. Only when the unphysical content of the spectral dimension
plays a particularly significant role should the new notion differ significantly from the spectral dimension.
In searching for such a new notion we took as guidance the observation reported in  recent
studies  \cite{Husain:2013zda, Santos:2015sva, Nozari:2015iba} (see also \cite{Witten} for earlier related proposals)
that in some instances the Stefan-Boltzmann law gives indications on the dimensionality of spacetime that
are consistent with the spectral dimension. One can view the Stefan-Boltzmann law as an indicator of spacetime dimensionality
since for a gas of radiation in a classical spacetime with $D+1$ dimensions the Stefan-Boltzmann law takes the form
\be
U\propto T^{D+1}. \label{eq:SB}
\ee
Actually several thermodynamical
relations are sensitive to the dimensionality
of spacetime, another example being the equation of state parameter $w\equiv P/\rho$, relating pressure $P$ and energy density $\rho$,
which for radiation in a classical spacetime with $D+1$ dimensions takes the form
\be
w=\frac{1}{D}.\label{eq:w}
\ee
These observations inspire our proposal of assigning a ``thermal dimension" to a quantum spacetime. Our recipe
involves studying the thermodynamical properties of radiation with on-shellness characterized by the (deformed) d'Alembertian
 of the relevant quantum-spacetime theory (the same deformed d'Alembertian used when evaluating the spectral dimension, but in its Lorentzian form). By looking at the resulting Stefan-Boltzmann law and equation of state one can infer the effective dimensionality of the
  relevant quantum spacetime.
  This notion of dimensionality has the advantage of being directly observable, a genuine physical property of the quantum spacetime,
  and, as we shall here show, fixes the shortcomings of the spectral dimension, while agreeing with it in
  some particularly noteworthy cases.

\section{Application to generalized Horava-Lifshitz scenarios.}

We start the quantitative part of our study by considering a class of generalized Horava-Lifshitz scenarios,
which has been the most active area of research on dynamical dimensional reduction \cite{Horava:2009if, Sotiriou:2011aa, DSRRSD}.
 These are cases where the momentum-space representation of the deformed d'Alembertian takes the form
\begin{equation}
\Omega_{\gamma_t \gamma_x}(E,p)= E^{2}-p^{2}+\ell_{t}^{2\gamma_{t}}E^{2(1+\gamma_{t})}-\ell_{x}^{2\gamma_{x}}p^{2(1+\gamma_{x})}\, .\label{eq:Omegagammatgammax}
\end{equation}
where $E$ is the energy, $p$ is the modulus of the spatial momentum, $\gamma_t$ and $\gamma_x$ are dimensionless parameters, and $\ell_t$ and $\ell_x$ are parameters with dimension of length (usually assumed to be of the order of the Planck length).

For this model it is known \cite{DSRRSD,Sotiriou:2011aa} that the UV value of the spectral dimension, obtained from the Euclidean version of the above d'Alembertian ($E^{2}+p^{2}+\ell_{t}^{2\gamma_{t}}E^{2(1+\gamma_{t})}+\ell_{x}^{2\gamma_{x}}p^{2(1+\gamma_{x})}$),
is
\begin{equation}
d_{S}(0)=\frac{1}{1+\gamma_t}+\frac{D}{1+\gamma_x}
\, .
\label{eq:dSLIV}
\end{equation}

In deriving the thermal dimension for this case we start from the logarithm of
the thermodynamical partition function \cite{Huang}, written so that the integration is explicitly taken over the full energy-momentum space:
\bea
\log Q_{\gamma_t \gamma_x}&=&-\frac{2V}{(2\pi)^{3}}\int dE \,d^{3}p \,\Big[ \delta(\Omega_{\gamma_t \gamma_x}) \, \Theta(E) \, \cdot \nonumber\\
&& \cdot \,2E\log\left(1-e^{-\beta E}\right)\Big]\,.\label{eq:LIVpartitionfunction}
\eea

Here $\beta$ is related to the Boltzmann constant $k_{B}$ and temperature via $\beta=\frac{1}{k_{B} T}$, and the delta function $\delta(\Omega_{\gamma_t \gamma_x})$ enforces the on-shell relation $\Omega_{\gamma_t \gamma_x}=0$.

\begin{figure}[h]
\centering
\scalebox{0.65}{\includegraphics{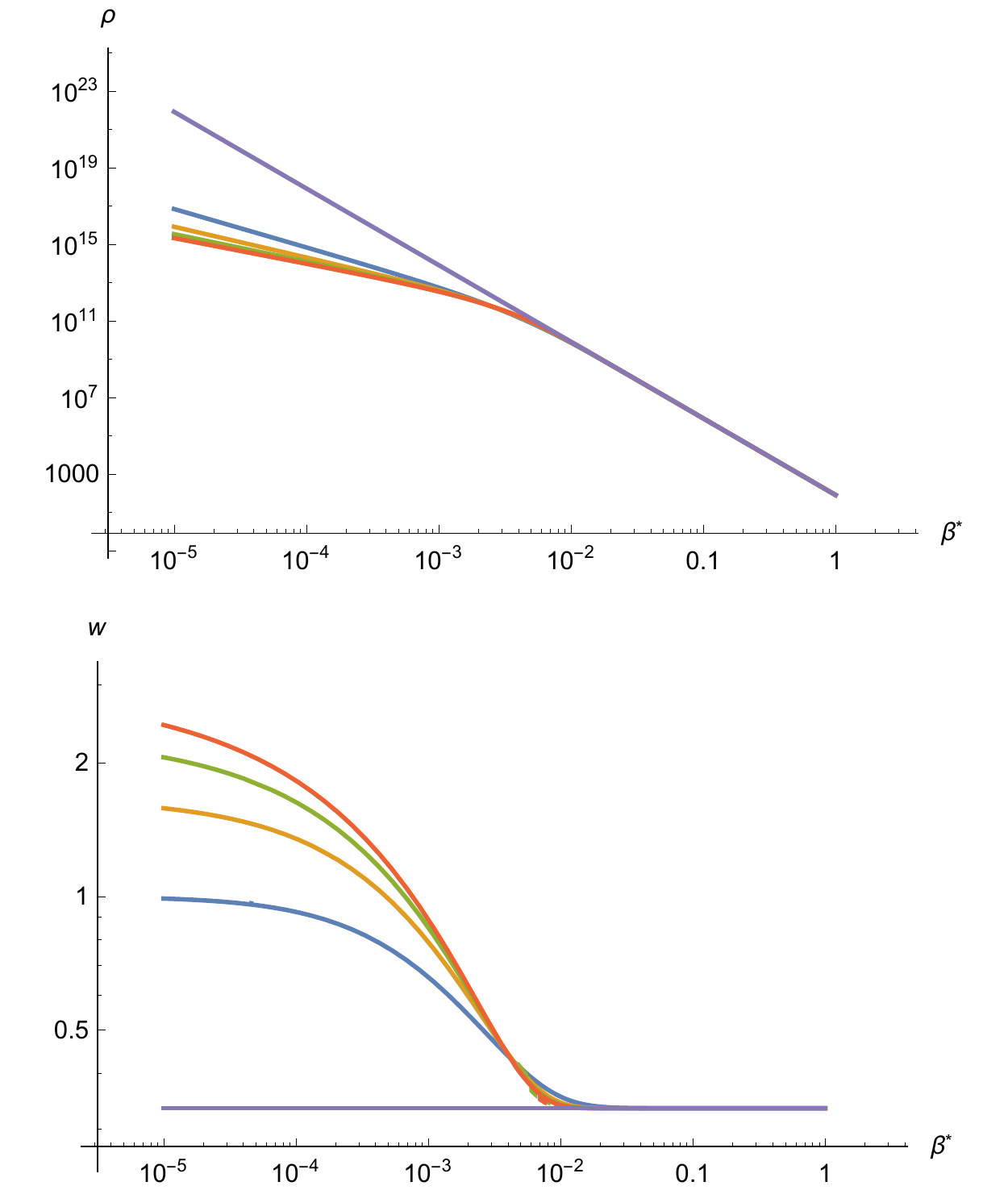}}
\caption{\label{fig:EOSLIV} Behavior of the energy density $\rho$ in arbitrary units (top panel) and of the equation of state parameter $w$  (bottom panel) as a function of $\beta^{*}\equiv 10^{-3} \beta k_{B}T_{P} $, according to the partition function $Q_{\gamma_t \gamma_x}$, for $\gamma_{t}=0$ and $\gamma_{x}=2$ (blue), $\gamma_{x}=4$ (orange), $\gamma_{x}=6$ (green), $\gamma_{x}=8$ (red). The purple line is the standard case, $\rho\propto T^{4}$ (top panel) and $w = 1/3$ (bottom panel). }
\end{figure}

From (\ref{eq:LIVpartitionfunction}) one obtains the energy density and the pressure respectively as
\be
\rho_{\gamma_t \gamma_x}\equiv -\frac{1}{V}\frac{\partial}{\partial \beta}\log Q_{\gamma_t \gamma_x}\,\,,
%\label{eq:rhodef}
%\ee
%\be
\,\,\,\,\,\,
p_{\gamma_t \gamma_x} \equiv \frac{1}{\beta}\frac{\partial }{\partial V}\log Q_{\gamma_t \gamma_x} \,.\label{eq:pdef}
\ee
In Figure \ref{fig:EOSLIV}  we show (for a few choices of $\gamma_{x} , \gamma_{t}$)
the resulting temperature dependence for the energy density and
 for the equation of state parameter.
For the UV/high-temperature values of $\rho_{\gamma_t \gamma_x}$
and $w_{\gamma_t \gamma_x}$ one can easily establish the following behaviors
 at high temperature, in agreement with the content
of Figure \ref{fig:EOSLIV}
\be
\rho_{\gamma_t \gamma_x} \propto T^{1+3\frac{1+\gamma_{t}}{1+\gamma_{x}}} \,\, ,
\,\,\,\,\,\,
w_{\gamma_t \gamma_x}=\frac{1+\gamma_{x}}{3(1+\gamma_{t})}\,.\label{eq:EOSLIV}
\ee

By comparison to (\ref{eq:SB}) and (\ref{eq:w}) one sees that both of these results
 give a consistent prediction for the ``thermal dimension" at high temperature, which is
 \be
d_{T}=1+3\frac{1+\gamma_{t}}{1+\gamma_{x}}\,.
\ee
Interestingly, in this case of
generalized Horava-Lifshitz scenarios
the thermal dimension agrees with spectral dimension, eq. (\ref{eq:dSLIV}), for $\gamma_{t}=0$,
but differs from the spectral dimension when $\gamma_{t}\neq 0$.

\section{Implications of active diffeomorphisms on momentum space.}

Generalized Horava-Lifshitz scenarios also give us an easy opportunity for comparing the properties
of the thermal dimension and of the spectral dimension under active diffeomorphisms on momentum space.
From this perspective the analysis is particularly simple for the case $\gamma_{x}=0, \gamma_{t}=1$, where
one has
\be
\Omega_{1,0}(E,p)= E^{2}-p^{2}+\ell_{t}^{2}E^{4}\, .
\ee
In light of the results reviewed and derived above we know that in this case
 the UV spectral dimension is $d_{S}=3.5$, while the UV thermal dimension is $d_{T}=7$.

Let us then contemplate a simple diffeomorphism on momentum space, the following
reparameterization of the energy variable: $E \rightarrow \tilde E = \sqrt{E^{2}+\ell_{t}^{2}E^{4}}$.
In terms of $\tilde E$ the d'Alembertian takes the standard special-relativistic form, $\Omega_{1,0}= \tilde E^{2}-p^{2}$, while the momentum space measure becomes non-trivial:
\be
d\mu(\tilde E,p)= \frac{d\tilde E d p\,\sqrt{2}\ell_{t} p^{2}\tilde E }{\sqrt{(1+4 \ell_{t}^{2}\tilde E^{2})(-1+\sqrt{1+4\ell_{t}^{2}\tilde E^{2}})}}
\label{newlabel}
\ee

When the above diffeomorphism on momentum space is an active one, the laws of physics are not invariant.  This is indeed what is found when comparing the thermodynamical properties of the ``$\tilde E,p$ theory" with d'Alembertian $\tilde E^{2}-p^{2}$ and momentum-space integration measure (\ref{newlabel}) and the ``$E,p$ theory" with (deformed) d'Alembertian  $\Omega_{1,0}(E,p)= E^{2}-p^{2}+\ell_{t}^{2}E^{4}$
and integration measure $dE \,d^{3}p$. In the ``$\tilde E,p$ theory" the logarithm of the thermodynamical partition function is
\bea
\log \tilde Q_{act.}&=&-\frac{2V}{(2\pi)^{3}}\int d \mu(\tilde E, p)\Big[\delta(\tilde E^{2}-p^{2})\Theta(\tilde E) \cdot \nonumber\\
&&\cdot 2\tilde E\log\left(1-e^{-\beta \tilde E}\right)\Big]\, \neq \log Q. \label{eq:activePartitionFunction}
\eea
Of course ultimately this leads us to obtain different values for the thermal dimension of these two theories. In fact, from the partition function (\ref{eq:activePartitionFunction}) one can easily find that at high temperatures the energy density behaves as $\rho\sim T^{3.5}$, while the equation of
state parameter is $w=0.4$. These numbers  point at a value of the UV thermal dimension of $d_{T}=3.5$.
Note that this result is different from the one that would follow from a passive diffeomorphism. In this case, the partition function in the $\tilde E,p$ variables would be straightforwardly obtained by a change of variables in eq. (\ref{eq:LIVpartitionfunction}):
\bea
\log \tilde Q_{pass.}&=&-\frac{2V}{(2\pi)^{3}}\int d \mu(\tilde E, p)\Big[\delta(\tilde E^{2}-p^{2}) \cdot \nonumber\\
&&\Theta(E(\tilde E)) 2E(\tilde E)\log\left(1-e^{-\beta E(\tilde E)}\right)\Big]\,\nonumber\\
&=& \log Q\,.
\eea
A passive diffeomorphism just relabels the same physical picture and of course the thermal dimension is not affected.

On the other hand, it can be easily seen that the spectral dimension is
not only invariant under passive diffeomorphisms but also under active diffeomorphisms on momentum space. In fact, active and passive diffeomorphisms have the same effect on the return probability $P(s)$ (eq. (\ref{eq:returnP})), that  of changing the integration variable (without changing the integral). Therefore the ``$\tilde E,p$ theory"
has the same UV spectral dimension ($d_{S}=3.5$)
as the ``$E,p$ theory".

%the thermal dimension of these two theories.

%The laws of physics are not invariant under such an active diffeomorphism
%on momentum space, and yet the spectral dimension is invariant: as evident
%from the form of (\ref{dsof0}) and (\ref{eq:returnP}),
%the ``$\tilde E,p$ theory" with d'Alembertian $\tilde E^{2}-p^{2}$ and momentum-space integration measure (\ref{newlabel})
%has the same UV spectral dimension ($d_{S}=3.5$)
%as the ``$E,p$ theory" with (deformed) d'Alembertian $\Omega_{1,0}(E,p)= E^{2}-p^{2}+\ell_{t}^{2}E^{4}$
%and integration measure $dE \,d^{3}p$.
%
%The fact that the ``$\tilde E,p$ theory" and the ``$E,p$ theory" are physically different is particularly clear
%when looking at thermodynamical properties, and of course ultimately this leads us to obtain different values for
%the thermal dimension of these two theories.
%In the ``$\tilde E,p$ theory" the logarithm of the thermodynamical partition function is
%\bea
%\log \tilde Q&=&-\frac{2V}{(2\pi)^{3}}\int d \mu(\tilde E, p)\Big[\delta(\tilde E^{2}-p^{2})\Theta(\tilde E) \cdot \nonumber\\
%&&\cdot 2\tilde E\log\left(1-e^{-\beta \tilde E}\right)\Big]\,.
%\eea
%From this one can easily find that at high temperatures the energy density behaves as $\rho\sim T^{3.5}$, while the equation of
%state parameter is $w=0.4$. These values  point at a value of the UV thermal dimension of $d_{T}=3.5$.
%
In summary, one finds that the UV spectral dimension of both the  ``$\tilde E,p$ theory" and the ``$E,p$ theory"
is 3.5, and 3.5 is also the value of the thermal dimension of the  ``$\tilde E,p$ theory",
but the ``$E,p$ theory" has UV thermal dimension of 7. It should be evidently seen as advantageous for
the thermal dimension\footnote{Some of us had contemplated in previous work \cite{DSRRSD, Amelino-Camelia:2013gna, Arzano:2015gda} the possibility of describing
the dimension of a quantum spacetime in terms of the duality with momentum space, by resorting to the ``Hausdorff dimension
of momentum space". However, at least as formulated in \cite{DSRRSD, Amelino-Camelia:2013gna, Arzano:2015gda}, that notion is only applicable to theories of the type
 of the ``$\tilde E,p$ theory", {\it i.e.} with undeformed d'Alembertian  (but possibly deformed measure of integration on
 momentum space).}
the fact that it assigns different UV dimension to the two very different ``$E,p$ theory"
and ``$\tilde E,p$ theory".

\section{Application to $f(E^{2} \! - \! p^{2})$ scenarios.}

Another scenario of significant interest is the one where the d'Alembertian is deformed into a function of itself:
$E^{2} \! - \! p^{2} \rightarrow f(E^{2} \! - \! p^{2})$. The structure of this scenario is very valuable for our purposes,
 but it also has intrinsic interest since it has been proposed on the basis of studies
of the Asymptotic-Safety approach \cite{Percacci:2007sz} and of the approach based on Causal Sets \cite{Belenchia:2014fda}.
We focus on a case which might deserve special interest from the quantum-gravity perspective,
 ass stressed in Ref.\cite{Percacci:2007sz}, such that the deformed d'Alembertian
takes the form
\be
\Omega_{\gamma}(E,p)=E^{2}-p^{2}-\ell^{2\gamma}\left(E^{2}-p^{2}\right)^{1+\gamma}\,,
\label{jocextra}
\ee
where the parameter $\gamma$ takes integer positive values and $\ell$ is a parameter with dimension of length.

For this case one easily finds\footnote{This elementary computation is reported in Ref. \cite{FrancescoPhD}.} that the UV spectral dimension is
\be
d_{S}(0)=\frac{4}{1+\gamma}\, ,
\ee
but the fact that this notion of the UV dimensionality of spacetime depends on $\gamma$ is puzzling and
points very clearly to the type of inadequacies of the spectral dimension that we are here concerned with.
In fact, in the UV limit the parameter $\gamma$ has no implications for the on-shell/physical properties of the (massless) theory.
  In general, massless particles governed by $\Omega_{\gamma}$
will be on-shell only either when
$$E^{2}=p^{2}$$
or when
$$E^{2}=p^{2}+\frac{1}{\ell^{2}}\,,$$
independently of the value of $\gamma$.
At low energies only $E^{2}=p^{2}$ is viable. For energies such that $E \geq 1/\ell$ also the second
possibility, $E^{2}=p^{2}+\frac{1}{\ell^{2}}$,
becomes viable. However, in the UV limit the two possibilities become indistinguishable, all particles
are governed by $E \simeq p$ just like in any 4-dimensional spacetime, because as $E \rightarrow \infty$
one has that $p^{2}+\frac{1}{\ell^{2}} \simeq p^2$. So without any need to resort  to complicated analyses we know
that this theory in the UV limit must behave like a 4-dimensional theory, in contradiction with the mentioned
result  for the UV spectral dimension.

\begin{figure}[h!]
\centering
\scalebox{0.65}{\includegraphics{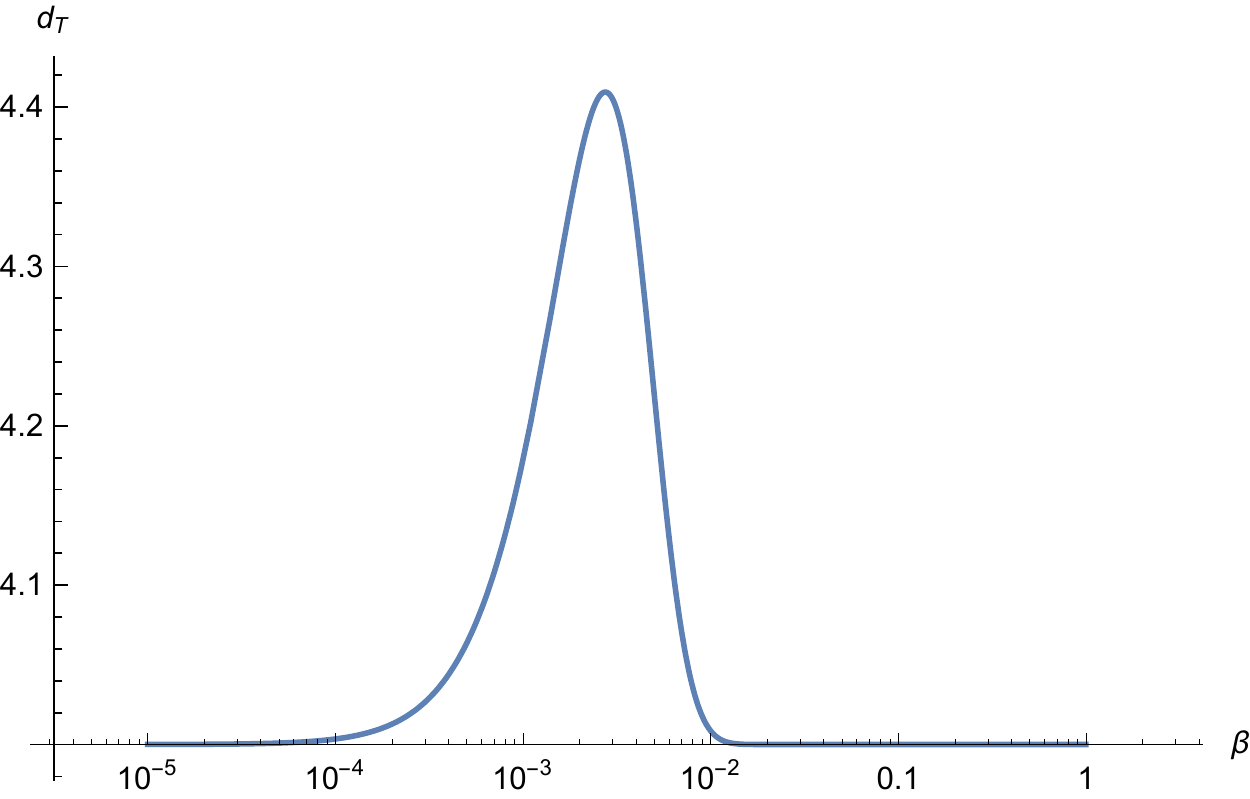}}
\caption{{ \label{fig:ASdT} Behaviour of the thermal dimension $d_{T}$ as a function of $\beta$. The thermal dimension is computed as  $d_{T}=1+\frac{1}{w}$, where the equation of state parameter $w$ is the one associated to the partition function $\log Q_\gamma$, with $\gamma=1$.  $\beta$ is in units of $10^{3}\beta_{P}$ (where $\beta_{P}=\frac{1}{k_{B}T_{P}}$ and $T_{P}$ is the Planck temperature).}}
\end{figure}

The UV value of our ``thermal dimension" is correctly 4, independently of $\gamma$.  This is easily seen by taking into account the deformation of d'Alembertian present in the $\Omega_{\gamma}$ of (\ref{jocextra}) for the analysis of
the partition function:
\be
\log Q_\gamma=-\frac{2V}{(2\pi)^{3}}\int dE d^{3}p \delta(\Omega_{\gamma})\Theta(E)2E\log\left(1-e^{-\beta E}\right),\label{eq:ASpartitionfunction}
\ee
Using the fact that
\be
\delta(\Omega_{\gamma})=\frac{\delta(E-p)}{2 p}+\frac{\delta(E-\sqrt{p^{2}+\frac{1}{\ell^{2}}})}{2\gamma\sqrt{p^{2}+\frac{1}{\ell^{2}}}}\,.\label{eq:gammadelta}
\ee
one easily finds that the UV behavior of thermodynamical quantities which is relevant to determine the thermal dimension is independent of $\gamma$, and in particular in
the UV the Stefan-Boltzmann law and the equation-of-state parameter take the form known for a standard 4-dimensional spacetime:
\be
\rho\propto T^{4} \,\, , \,\,\,\,
w=\frac{1}{3} \, .
\ee
So indeed in this scenario the UV value of the thermal dimension is 4. The theory
does have ``dynamical running of the dimensionality of spacetime" in a regime where the temperature is close to the Planckian temperature,
as one should expect on the  basis
of  the fact that the parameter $\gamma$ does have a role in  the theory
when the energy is greater than $1/\ell$, as long as the energy is not
 big enough to render $p^{2}$ and $p^{2}+\frac{1}{\ell^{2}}$ indistinguishable.
 This is shown in figure \ref{fig:ASdT}, where we plot the thermal dimension (inferred from the behaviour of the equation of state parameter and from the running of the energy density with temperature) as a function of $\beta$.

We attribute the disastrous failures of the spectral dimension in this case to a combination of its sensitivity to off-shell properties
and its reliance on the Euclidean d'Alembertian. It is noteworthy that for the Euclidean d'Alembertian\footnote{Note that in order to have the Euclidean version of the d'Alembertian $\Omega_{\gamma}(E,p)$ one has to  Wick-rotate also the parameter $\ell$ \cite{Calcagni:2014wba}.},
%  { \bf the fact that  the part in $\gamma$ has opposite sign implies in turn that  the second solution to the mass-shell constraint would be $E^{2}=p^{2}-\frac{1}{\ell^{2}}$, which  gives imaginary energy in the rest frame}
\be
\Omega^{[Euclidean]}_{\gamma}=E^{2}+p^{2} +\ell^{2\gamma} (E^{2}+p^{2})^{1+\gamma}\, ,
\ee
in the UV limit one can neglect $E^{2}+p^{2}$ with respect to $\ell^{2\gamma} (E^{2}+p^{2})^{1+\gamma}$. Instead for on-shell modes
 of the original
Lorentzian $\Omega_{\gamma}$
one can never neglect $E^{2}-p^{2}$ with respect to $\ell^{2\gamma} (E^{2}-p^{2})^{1+\gamma}$.

\section{Closing remarks.}

The exciting realization that the UV dimension
of spacetime might be different from its IR dimension adds significance to the
old challenge of describing the dimension of a quantum spacetime.
We here argued that it is crucial to link this issue to observable properties. After all what we mean
in physics by ``dimension of spacetime" must inevitably be something we measure. Moreover, only
by relying on a truly physical/observable characterization are we assured to compare different
theories in conclusive manner.

We here exposed fully the inadequacy of the spectral dimension for these purposes.
The fact that this notion involves an unphysical Euclideanization
could already lead to this conclusion. We feel that our observation about the undesirable invariance
of the spectral dimension under active diffeomorphisms of momentum
space should casts another shadow on the usefulness of the spectral dimension. The fact that
one obtains different spectral dimensions
for alternative formulations of the same
physical theory (formulations that differ only for what concerns unphysical off-shell modes)
 should leave no residual doubts.

We here proposed a notion of dimensionality which is free from the shortcomings of the spectral dimension, since it relies on the analysis of observable thermodynamical properties of radiation in the quantum spacetime.
Only experience with its use will gradually say if our notion of thermal dimension
of a quantum spacetime is not only physical but also particularly useful. We conjecture
that it will prove to be very valuable at least for studies of the early universe, which is anyway the context where the UV dimension of spacetime should find its
most significant applications \cite{Amelino-Camelia:2013tla, Amelino-Camelia:2015dqa}.

\section*{Acknowledgements.} GAC and GG acknowledge support from the John Templeton Foundation. GS acknowledges the support by the CAPES-ICRANet program financed by CAPES - Brazilian Federal Agency for Support and Evaluation of Graduate Education within the Ministry of Education of Brazil through the grant BEX 13955/13-6.

\end{document}